\documentclass[aps,prl,twocolumn,preprintnumbers,linenumber,amsmath,amssymb,superscriptaddress]{revtex4-1}
\usepackage{graphicx}
\usepackage{subfigure}
\usepackage{epsfig}
\usepackage{dcolumn}
\usepackage{bm}
\usepackage{dcolumn}
\usepackage{color}
\usepackage{amsbsy}
\usepackage{lineno}
\usepackage{blindtext}

\def\avg#1{\left\langle#1\right\rangle}
\def\bra#1{\left\langle#1\right|}
\def\ket#1{\left|#1\right\rangle}

\def\be{\begin{equation}}       \def\ee{\end{equation}}
\def\bea{\begin{eqnarray}}      \def\eea{\end{eqnarray}}
\def\ba{\begin{array} }
\def\ea{\end{array} }
\def\bnum{\begin{enumerate} }
\def\enum{\end{enumerate}}

\def\nn{\nonumber}
\def\pa{\partial}
\def\=>{\Rightarrow}
\def\>{\rightarrow}
\def\A{\uparrow}
\def\V{\downarrow}

\def\eye2{Fathbb{I}}

\def\Eq#1{Eq.~(\ref{#1})}
\def\Fig#1{Fig.~\ref{#1}}

\renewcommand{\>}{\rangle}

\begin{document}
\title{Edge stability and edge quantum criticality in 2D interacting topological insulators}
\author{Zi-Xiang Li}
\affiliation{Institute for Advanced Study, Tsinghua University, Beijing 100084, China}
\affiliation{Department of Physics, University of California, Berkeley, CA, 94720, USA}
\author{Hong Yao}
\email{yaohong@tsinghua.edu.cn}
\affiliation{Institute for Advanced Study, Tsinghua University, Beijing 100084, China}

\begin{abstract}
 Robustness of helical edge states in 2D topological insulators (TI) against strong interactions remains an intriguing issue. Here, by performing the first \textit{sign-free} quantum Monte Carlo (QMC) simulation of the Kane-Mele-Hubbard-Rashba model which describes an interacting 2D TI with \textit{two-particle backscattering} on edges, we verify that the gapless helical edge states are robust against a \textit{finite} range of two-particle backscattering when the Coulomb repulsion is not strong. However, when the Coulomb repulsion is strong enough, the helical edge states can be gapped by \textit{infinitesimal} two-particle backscattering, resulting in edge magnetic order. We further reveal universal properties of the magnetic edge quantum critical point (EQCP). At magnetic domain walls on edges, we find that a fractionalized charge of \textit{e}/{2} emerges. Implications of our results to recent transport experiments in the InAs/GaSb quantum well, which is a 2D TI with strong interactions, will also be discussed.
\end{abstract}
\date{\today}
\maketitle

{\bf}

Topology has played an increasingly important role in condensed matter physics in the past few decades \cite{XGWenbook,Fradkinbook}, especially after discoveries of quantum Hall effect \cite{TKNN,Laughlin1983}, high-temperature superconductors \cite{Anderson1987,Kivelson1987,Xiaogang-2006RMP}, and recently topological insulators \cite{Hasan-Kane-10,Qi-Zhang-11}. Especially, tremendous progress has been made in both theoretical and experimental understandings of topological insulators (TI) with negligible or weak interactions \cite{Kane2005a,Kane2005b,Bernevig2006prl,Bernevig2006science, Fu2007a,Moore2007,Roy2009, Qi2008,Schnyder2008, Kitaev2009, Molenkamp2007,Hasan2008,Chen2009}. For instance, 2D TIs are topological state of matter supporting the gapless helical edge states which are protected by time reversal (TR) symmetry when interactions are relatively weak. As strong interactions may destabilize the gapless excitations on boundaries  \cite{Fidkowski2010,Qi2013,Yao2013,Ryu2012,Ashvin2013prx,Wang2014prb,Wang2014science, DHLee2011,Xu2015prb,Ziani2015,Mudry2015,Gefen2017,Sangiovanni2017} of symmetry-protected topological phases \cite{Hasan-Kane-10,Qi-Zhang-11,ZCGu2009prb, Pollmann2010prb,Chen2011prb,Chen2012science,YMLu2012}, understanding the interplay between topology and interactions, especially in experimentally accessible topological materials, remains an intriguing problem.

2D TIs were first discovered in HgTe quantum wells (QWs) \cite{Bernevig2006science,Molenkamp2007} and later in InAs/GaSb QWs \cite{Liu2008,Du2011,Moler2014}. Although the correlation effect in the helical edge states of the HgTe QWs is relatively weak (a Luttinger liquid \cite{Haldane1981,Kane1992,Nagaosa1994,Giamarchibook,Wu2006,Xu2006} with edge Luttinger parameter $K\approx 0.81$ \cite{Joseph2009,Kane2009prb}, close to the noninteracting limit of $K=1$), recent transport experiments \cite{Du2015a,Du2015b} reported evidences of strong interactions in the helical edge states of the InAs/GaSb QWs (with edge Luttinger parameter $K\approx 0.22$ \cite{Du2015b}). Consequently, it is desired to thoroughly investigate the robustness of gapless helical edges states of 2D TIs with strong interactions due to its importance in both fundamental physics and future applications.

It was predicted that helical edge states are perturbatively stable against weak two-particle backscattering if the edge Luttinger parameter (which is mainly determined by local Coulomb interaction) $K$$>$$\frac12$ \cite{Wu2006,Xu2006}. When the local Coulomb interaction is strong enough such that $K$$<$$\frac12$, the two-particle backscattering interaction becomes relevant and an infinitesimal two-particle backscattering can lead to edge magnetic order, which breaks time-reversal (TR) symmetry spontaneously and opens up a gap in helical edge states \cite{Wu2006,Xu2006}. However, such perturbative analysis of weak two-particle backscattering may not be directly applied for the case of strong interactions. Consequently, non-perturbative methods are desired to explore the effect of strong interactions in topological insulators such as the InAs/GaSb QWs.

\begin{figure}
\subfigure[]{\includegraphics[height=3.1cm]{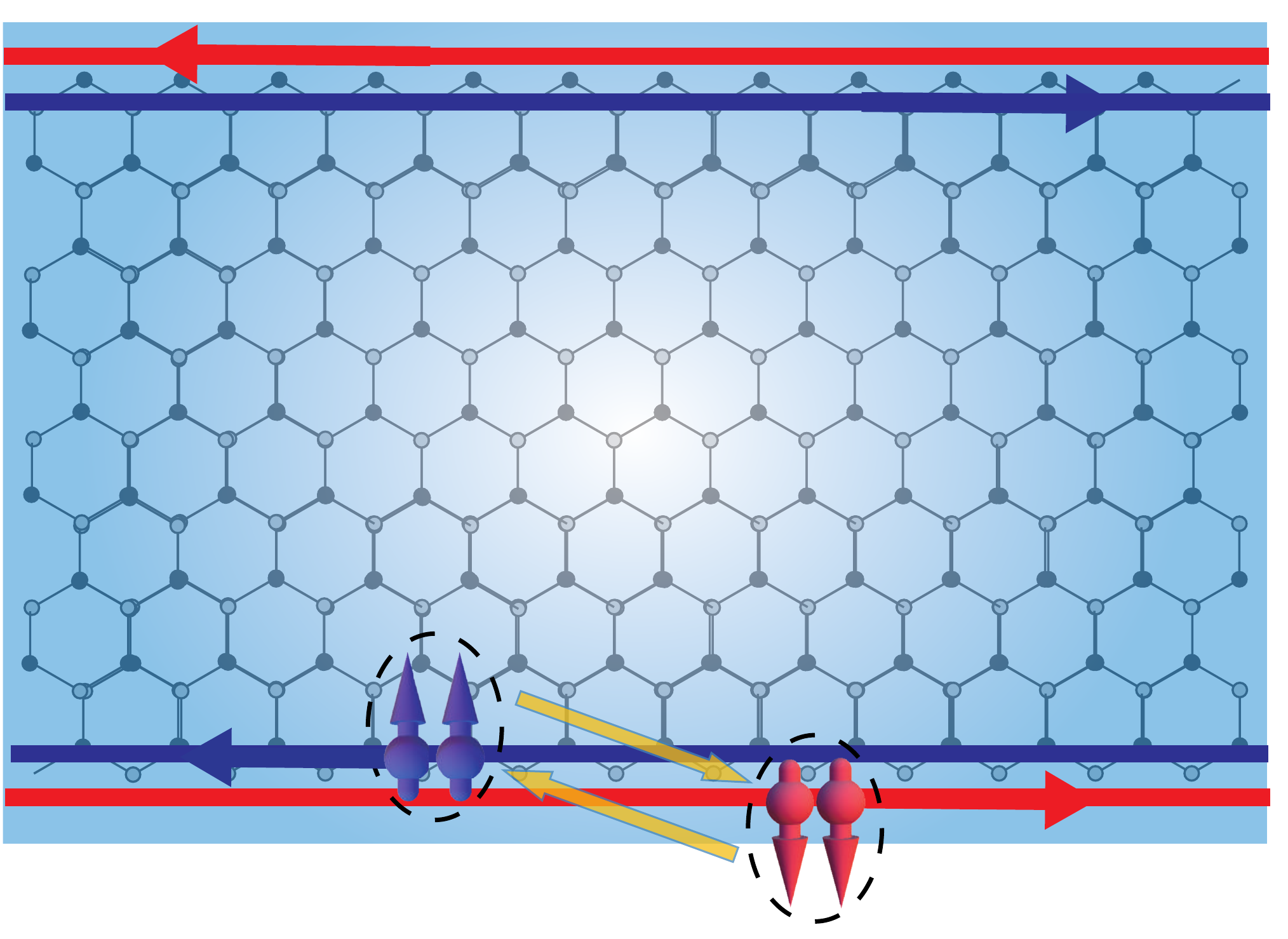}}
\subfigure[]{\includegraphics[height=3.1cm]{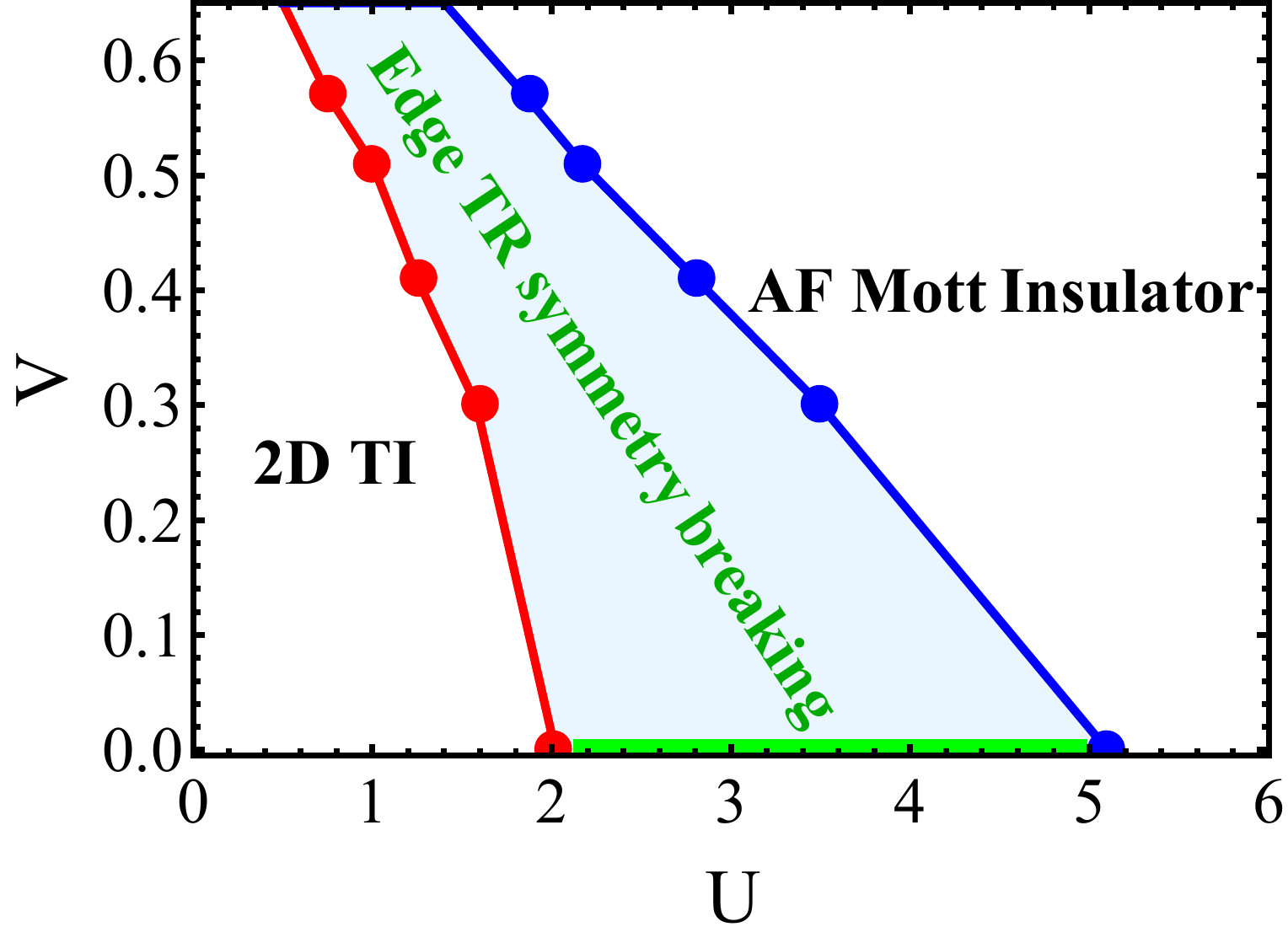}}
\caption{(a) Two-particle backscattering in helical edge states of an interacting 2D TI is allowed when the Rashba interaction is present. (b) The quantum phase diagram of the Kane-Mele-Hubbard-Rashba model allowing edge two-particle backscattering. $U$ represents Hubbard repulsion and $V$ labels Rashba interaction which induces two-particle backscattering on edges. The dots are data points obtained by our sign-free QMC simulations. The edge and bulks QCP are clearly separated; in the intermediate region only edges break the TR symmetry while the bulk is TR invariant.}
\label{fig1}
\end{figure}

To fill in this gap, by performing quantum Monte Carlo (QMC), we study the stability of helical edge states of 2D interacting TIs in the presence of both strong Coulomb (Hubbard) repulsion and Rashba interactions that cause two-particle backscattering on edges. Here, it is the Rashba interactions that break the $U$(1) spin-rotational symmetry and render a finite two-particle backscattering in helical edge states. Usual QMC simulations of fermionic quantum models often encounter the notorious fermion-sign-problem \cite{Loh1990,Troyer2005,Wu2005}, which prevents the application of conventional QMC \cite{BSS1981,Hirsch1985,Assaad2008,Berg2012} from accurately studying systems with large size and at low temperature. However, using the novel Majorana algorithm introduced by us \cite{LJY2015}, sign-free Majorana QMC simulations \cite{LJY2015,LJY2016,Xiang2016} can be performed on the 2D interacting TI model on the honeycomb lattice with both Hubbard repulsion and Rashba interaction [see \Eq{TI} below], which we call the Kane-Mele-Hubbard-Rashba (KMHR) model. Note that previous sign-free QMC can \textit{only} study 2D interacting TIs \textit{without} edge two-particle backscattering \cite{Wu2011,Assaad2011,Assaad2013} which cannot accurately address the issue of quantum critical behaviors of the edge magnetic ordering and that the effect of single-particle Rashba hopping term was studied by QMC in the presence of the sign problem \cite{Assaad2014prb} or by other methods which involve approximations \cite{Rachel2014}.

From our sign-free QMC simulations of the KMHR model, we have verified that the helical edge states of an interacting 2D TI are robust against a finite range of two-particle backscattering when the Hubbard interaction is relatively weak. We show that the edge states spontaneously break time-reversal symmetry when the Rashba interaction exceeds a critical value and obtain the critical exponents of the edge quantum critical point (EQCP) in 1+1 dimensions. Moreover, we find that a fractionalized charge of $e/2$ emerges at the magnetic domain wall on the edges when magnetic ordering occurs in the edges of the interacting TI. To the best of our knowledge, this is the first time that the effect of two-particle backscattering can be simulated numerically-exactly in 2D interacting TI by the unbiased and nonperturbative approach of sign-free QMC.

\textbf{The 2D interacting TI model:} We first introduce a minimal model on the honeycomb lattice describing a 2D interacting topological insulator allowing two-particle backscattering on edges:
\bea\label{TI}
&&H= -t \sum_{\avg{ij}\sigma}c^\dagger_{i\sigma}c_{j\sigma} + \sum_{\avg{\avg{ij}}}(i\lambda \nu_{ij}\sigma^z_{\alpha\beta} c^\dagger_{i\alpha}c_{j\beta} + H.c.) \nonumber\\
&&~~~~~~+ U \sum_{i} n_{i\uparrow}n_{i\downarrow} + V \sum_{\avg{ij}} \big(c^\dagger_{i\uparrow}c^\dagger_{j\uparrow}c_{i\downarrow}c_{j\downarrow} + H.c.\big),
\eea
where $c^\dagger_{i\sigma}$ creates an electron on site $i$ with spin polarization $\sigma = \uparrow,\downarrow$, $n_{i\sigma} = c^\dagger_{i\sigma}c_{i\sigma}$ is the particle number operator. The first term in \Eq{TI} represents the usual nearest-neighbor hopping while the second one with $\nu_{ij}$=$\pm 1$ is the Kane-Mele spin-orbital coupling (SOC) \cite{Kane2005a}. The Kane-Mele SOC reduces the spin rotational symmetry from $SU(2)$ to $U(1)$. The Hubbard $U$ term mimics the effects of Coulomb repulsion. The Hamiltonian in \Eq{TI} without the last term is called Kane-Mele-Hubbard (KMH) model, for which a sufficiently strong $U$ would favor AF ordering with magnetic moments in the $xy$ plane\cite{Rachel2010,JXLi2011,Wu2011,Assaad2011,Assaad2013, Sangiovanni2015}. For the $V$ term in \Eq{TI}, we call it  ``Rashba interaction'' because this two-particle spin-flip term can be induced by the Rashba SOC. This is why the Hamiltonian in \Eq{TI} is denoted as the ``Kane-Mele-Hubbard-Rashba'' model. The Rashba interaction only respects a residual $Z_2$ spin symmetry; namely $(-1)^{N_\A}$ is conserved where $N_\A$ is the number of spin-up electrons. In other words, a finite Rashba interaction $V$ further reduces the spin symmetry from $U(1)$ to $Z_2$, which introduces two-particle spin-flip backscattering in the helical edge states. Note that the single-particle backscattering in helical edge states is forbidden even in the presence of finite Rashba interaction due to the TR symmetry.

Among all the two-particle scattering processes respecting the TR symmetry, only the two-particle spin-flip backscattering has the potential to open up a gap in the helical edge states. Consequently, this Rashba interaction is the leading term which can destabilize the gapless helical edge states of the 2D TI. It was thought for many years that this model with both Hubbard repulsion and spin-flip interaction cannot be simulated by sign-problem-free QMC \cite{Wu2011,Assaad2011,Assaad2013}. However, employing the Majorana algorithm recently introduced in Ref. \cite{LJY2015}, we show that the KMHR model in \Eq{TI} can be simulated by QMC without encountering the notorious fermion sign problem such that accurate large-scale QMC simulations of the model can be performed to investigate the edge stability of the 2D interacting TI.

Here we briefly discuss how Majorana representation can help solve the sign problem of the KMHR model. First, we introduce the Majorana representation of spin-1/2 electrons: $c_{i\sigma} = \frac{1}{2}(\gamma^1_{i\sigma} + i \gamma^2_{i\sigma})$, $c^\dagger_{i\sigma} = \frac{1}{2}(\gamma^1_{i\sigma} - i\gamma^2_{i\sigma})$, where $\gamma^\tau_{i\sigma}$ are Majorana fermions operators with $\tau=1,2$ representing Majorana index and $\sigma=\uparrow,\downarrow$ spin index. In the Majorana representation, we perform Hubbard-Stratonovich transformations such that the decoupled Hamiltonian at imaginary time $\tau$ respects two anti-commuting anti-unitary symmetries: $T^- = i \sigma^y \tau^x K$ and $T^+ = \sigma^x \tau^x K$ (see the SM for details). According to the Majorana TR principle \cite{LJY2016,Xiang2016}, the QMC simulation using the Majorana algorithm is sign-free!

\begin{figure}[t]
\subfigure[]{\includegraphics[height=2.8cm]{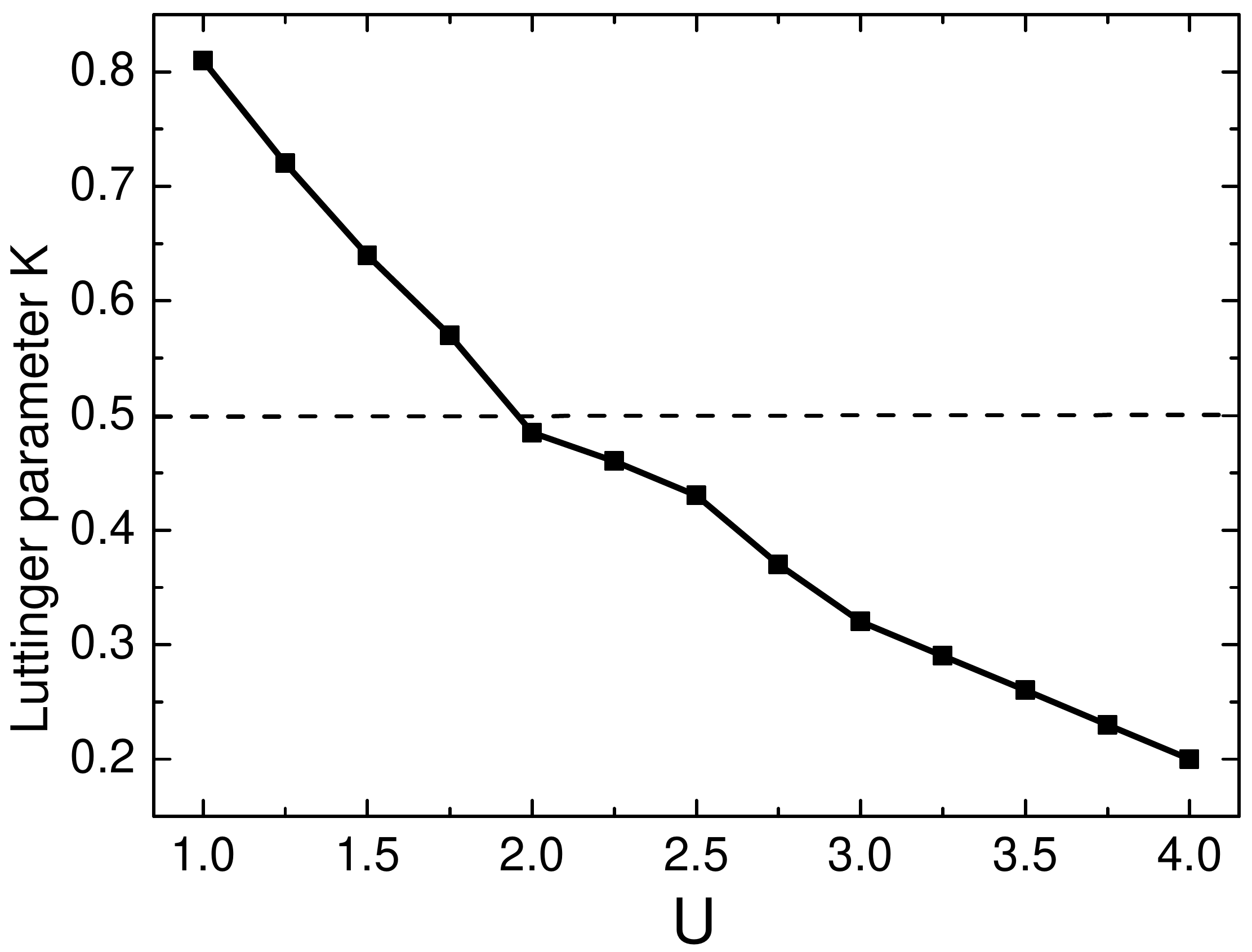}}
\subfigure[]{\includegraphics[height=2.8cm]{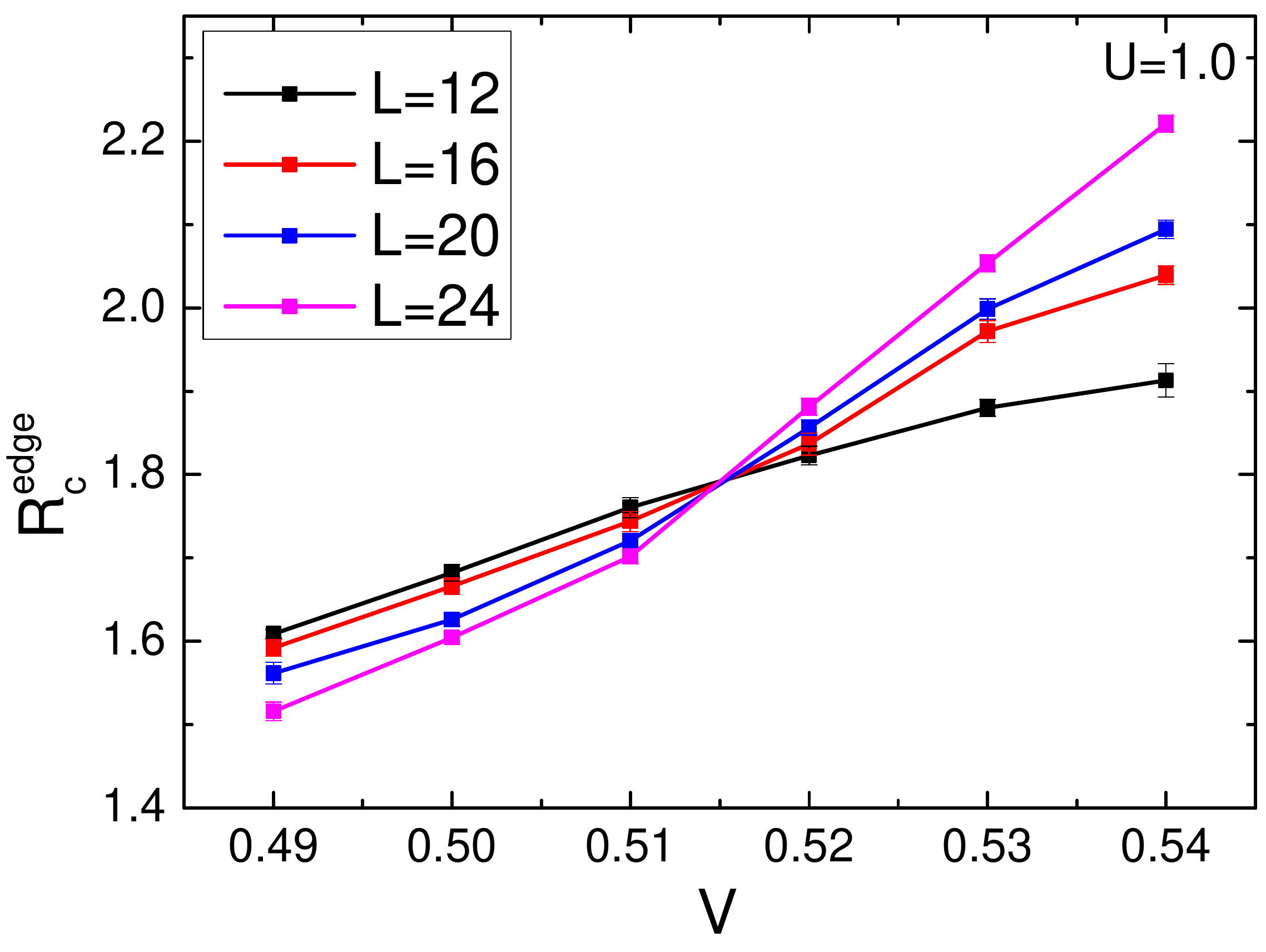}}
\subfigure[]{\includegraphics[height=2.8cm]{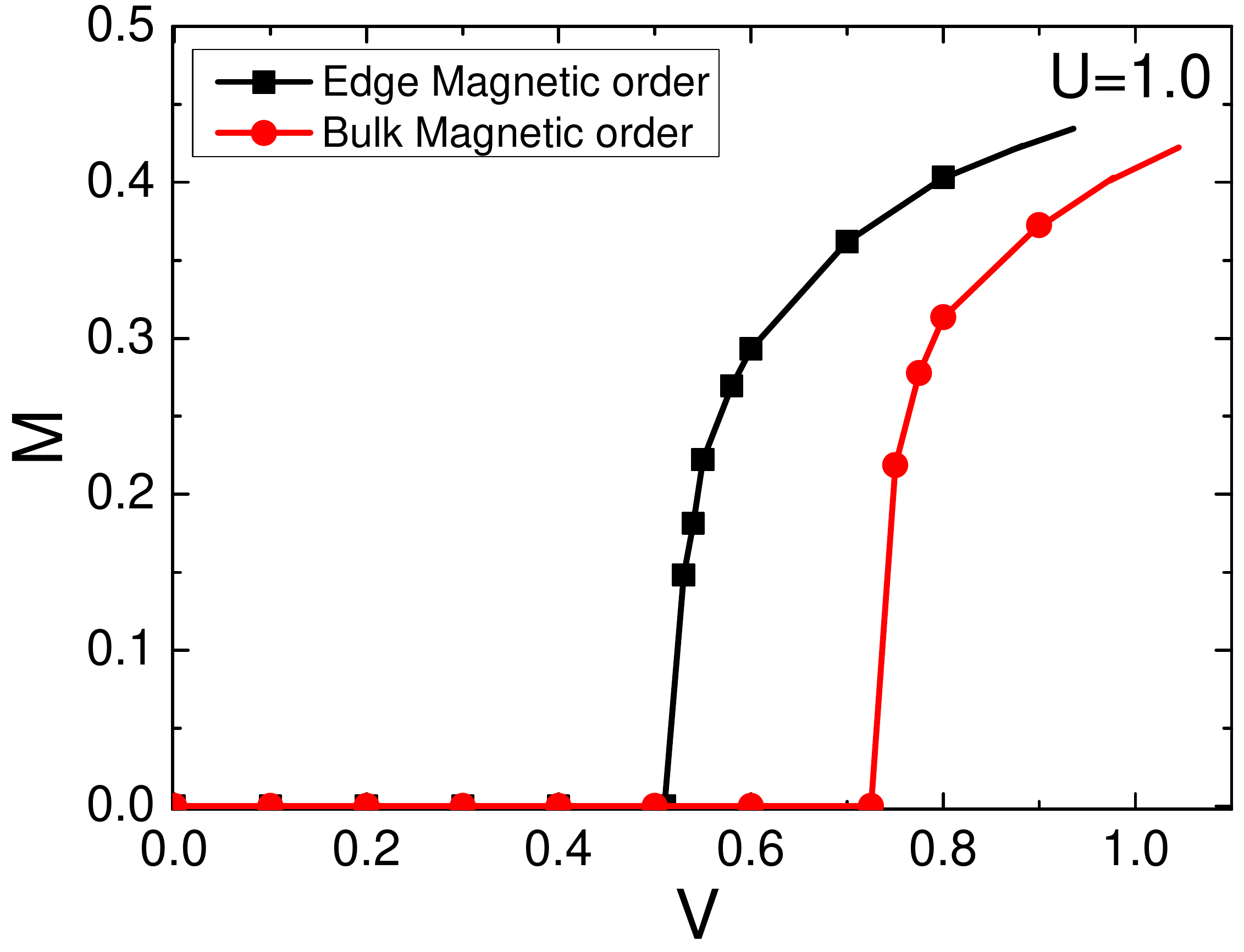}}~~~	
\subfigure[]{\includegraphics[height=2.8cm]{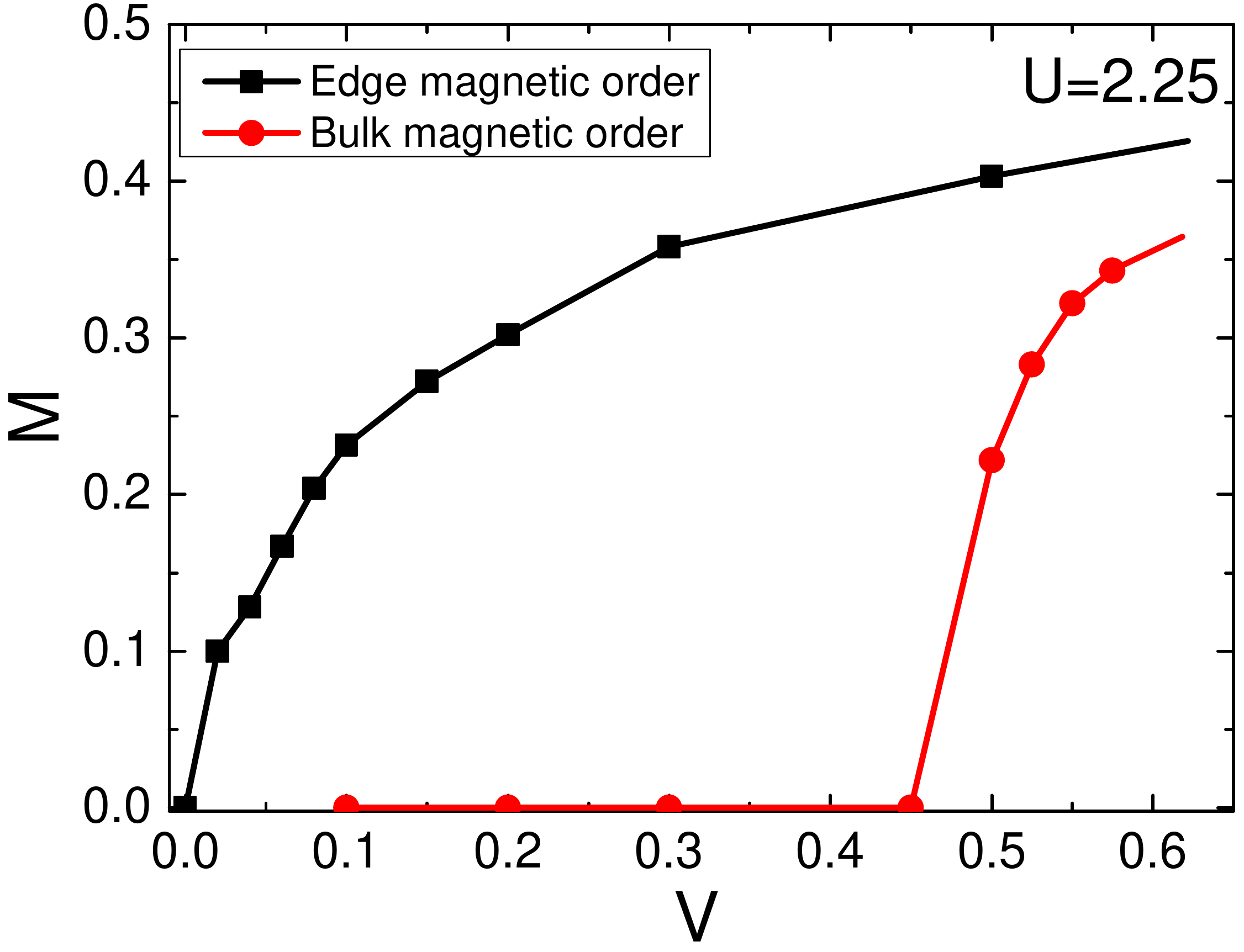}}							
\caption{(a) The QMC results of the Luttinger parameter $K$ as the function of $U$ while the spin-flip interaction $V$=0. (b) For $U$=1.0, the QMC results of edge RG-invariant ratio $R_c^\textrm{edge}$ as the function of $V$, from which we obtain the edge critical values of Rashba interaction $V^\textrm{edge}_c$$\approx $0.515. (c) The magnetic order parameter $\avg{S^x_i}$ at edge and bulk for $U=1.0t$ ( $K$$>$$\frac12$ in the limit of $V=0$) as a function of $V$. (d) The magnetic order parameter $\avg{S^x_i}$ at edge and bulk for $U=2.25t$ ( $K$$<$$\frac12$ in the limit of $V=0$) as a function of $V$. }
\label{fig2}
\end{figure}

\textbf{The quantum phase diagram:}
We perform large-scale sign-free projector QMC \cite{Sugiyama1986,Sorella1989,White1989} to study the ground state properties of the 2D interacting topological insulator with two-particle spin-flip scattering. In our simulations, we set $\lambda=0.1t$ unless noted otherwise. To investigate the stability of helical edge states with respect to two-particle spin-flip scattering, we consider the system with \textit{periodic} boundary condition (PBC) along the $x$-direction but \textit{open} boundary condition (OBC) along the $y$-direction, as shown in \Fig{fig1}(a). Since the bulk is fully gapped, we focus on the helical edge states, which can be effectively described by the LL theory \cite{Haldane1981,Kane1992,Giamarchibook} with the following Hamiltonian in the continuum:
\bea\label{fermionH}
&&H=\int dx \Big[v_F(\psi^\dag_{R\A}i\pa_x\psi_{R\A}-\psi^\dag_{L\V}i\pa_x\psi_{L\V}) \nn\\
&&~~~~~~+g_2 \psi^\dag_{R\A}\psi_{R\A}\psi^\dag_{L\V}\psi_{L\V}+g(\psi^\dag_{R\A}\psi^\dag_{R\A} \psi_{L\V}\psi_{L\V}+H.c.)\Big],~~~
\eea
where $\psi_{R\A}$ and $\psi_{L\V}$ are annihilation operators of spin-$\A$ right-moving and spin-$\V$ left-moving electrons on the edge, respectively. In \Eq{fermionH}, the last term represents $g(\psi^\dag_{R\A}(x)\psi^\dag_{R\A}(x+\delta)\psi_{L\V}(x)\psi_{L\V}(x+\delta)+H.c.)$, where $\delta$ is the short-distance cutoff. Here $v_F$ is the Fermi velocity of non-interacting electrons on edges and is approximately proportional to the Kane-Mele SOC $\lambda$. $g_2$ is the two-particle forward-scattering and is proportional to $U$; $g$ is the two-particle backscattering induced by the Rashba interaction $V$. Note that single-particle backscattering is forbidden by the TR symmetry even in the presence of Rashba SOC and that in \Eq{fermionH} we neglect the interactions $g_4[\psi^\dag_{R\A}(x)\psi_{R\A}(x)\psi^\dag_{R\A}(x+\delta)\psi_{R\A}(x+\delta) +R\!\to\!L]$ which only renormanizes the Fermi velocity.

The fermionic edge theory in \Eq{fermionH} can be bosonized with the following sine-Gordon Lagrangian density:
\bea
{\cal L}=\frac1{2\pi K} \left[\frac1v(\pa_\tau\phi)^2+v(\pa_x\phi)^2\right]-\frac{g}{(2\pi\delta)^2} \cos(4\phi),~~~
\eea
where $\phi$ is the boson field obtained from bosonizing the fermions on the edge \cite{Wu2006,Xu2006}, $v\!=\!v_F\sqrt{1-(\frac{g_2}{v_F})^2}$ is the renormalized velocity of bosons, $K=\sqrt{\frac{1-g_2/v_F}{1+g_2/v_F}}$ is the Luttinger parameter which decreases as $U$ increases, and $g$ characterizes the two-particle backscattering whose scaling dimension is $\Delta_g=4K$. Consequently, two-particle spin-flip backscattering is irrelevant when the Luttinger parameter $K$$>$$\frac12$ but relevant when $K$$<$$\frac12$ \cite{Kane1992,Wu2006,Xu2006}. We expect that the helical edge states are stable against infinitesimal two-particle backscattering when the Hubbard interaction $U$ is not sufficiently strong because of the irrelevance of weak two-particle backscattering for $K$$>$$\frac12$. When the two-particle backscattering is relevant for $K<\frac12$, the mass gap opens even for weak Rashba interaction $V$ and the gap is given by $\Delta\approx \frac1{\delta}g^{\frac{1}{2-4K}}$ for small $g$. This conclusion comes from the perturbative analysis of $g$. But, for $K>\frac12$, strong enough $g$ should also drive a mass gap. This can be obtained from the QMC calculations below.

We first compute the values of Luttinger parameter of the helical edge states for different Hubbard interaction $U$ while setting the Rashba interaction $V=0$. The edge Luttinger parameter $K$ can be extracted by measuring the correlation function of magnetic order parameter:
\bea\label{correlation}
C(r) = \frac{1}{L_x}\sum_{x_i=1}^{L_x} \avg{S^x_{i,A} S^x_{i+r,A}},
\eea
where $S^x_{i,A} = \frac{1}{2}(c^\dagger_{i\uparrow} c_{i\downarrow} + c^\dagger_{i\downarrow} c_{i\uparrow})$ are the spin operators in the $A$-sublattice of the unit cell $i=(x_i,y_i)$ on the edge $y_i=L_y$ ($L_x$ and $L_y$ are the number of unit cells along the $x$ and $y$ directions, respectively). Note that due to the U(1) spin-rotational symmetry at $V=0$, using $S^x$, $S^y$, or $\vec S\cdot \hat n$  with any $\hat n$ in \Eq{correlation} would give rise to the same value of $K$. In other words, we compute the correlation functions between edge sites of the $A$ sublattice, which are the tips on the top zigzag edge of the honeycomb lattice, as shown in \Fig{fig1}(a). In our QMC simulations, we choose $L_x=2L_y=L$. According to the LL theory, the magnetic correlation function in \Eq{correlation} scales as $C(r)\sim r^{-2K}$ for large $r$ . In order to reduce the finite-size effect, we extract the Luttinger parameter from fitting $C(r_\textrm{max}$$=$$\frac{L}{2})\sim L^{-2K}$, which is the correlation function at the largest possible separation between two sites along the edge of the lattice. The obtained values of $K$ as a function of $U$ is shown in \Fig{fig2}(a). It is clear that $K$ decreases as $U$ is increased, as expected. At $U=U^\textrm{edge}_{c} \approx 1.85t$, the Luttinger parameter $K=K_c=\frac12$. The bulk AF magnetic ordering can occur for $U>U^\text{bulk}_{c}$ where $U^\textrm{bulk}_{c}\approx 5.15t$.

From bosonization analysis, the helical edge states should be stable against a finite range of two-particle backscattering for $U$$<$$U^\textrm{edge}_{c}$. Thus, for $U$$<$$U^\textrm{edge}_c$, we expect that a sufficiently large Rashba interaction can destabilize the helical edge states by inducing magnetic order. To illustrate this, we set $U=1.0t$ (the corresponding $K\sim 0.81$ in the limit of $V=0$) and vary the Rashba interaction $V$ to study the edge instability against the two-particle backscattering. The Rashba interaction with $V>0$ would favor magnetic ordering in the $x$ direction more than the $y$ direction. We use the RG-invariant quantity $R^\textrm{edge}_c$ in terms of second-moment correlation length \cite{Assaad2015} of edge correlation function $C(r)$ to identify the EQCP ($R^\textrm{edge}_c$ is defined in the SM). At the putative EQCP, $R^\textrm{edge}_c$ should cross at the same point for different system sizes. From the results of RG-invariant quantity $R_c^\textrm{edge}$, as shown in \Fig{fig2}(b), we obtain the edge critical value $V^\textrm{edge}_c \approx 0.515t$. For $V$$>$$V^\textrm{edge}_c$, the edge ferromagnetic (FM) ordering spontaneously occurs (the magnetic moments on sites in the $A$ sublattice point to the same direction \cite{Wu2015}). When $V$ is further increased such that $V>V^\textrm{bulk}_c\approx 0.73t$, the bulk AF ordering occurs. The magnetic order parameter $M$$=$$\avg{S^x_i}$ for $U=1.0t$ as a function of $V$ is shown in \Fig{fig2}(c).

For $U^\textrm{edge}_{c}$$<$$U$$<$$U^\textrm{bulk}_{c}$ or equivalently $K$$<$$\frac12$, an infinitesimal two-particle backscattering on the edges can gap out helical edge states by inducing magnetic order \cite{Wu2006,Xu2006,Wu2011}. For instance, we set $U=2.25t$ and vary the Rashba interaction $V$ to study the edge instability. As shown in \Fig{fig2}(d), the edge magnetic ordering already occurs  for a very weak $V$, indicating that the two-particle backscattering is relevant and that edge magnetic order is induced as long as $V>V^\textrm{edge}_c$$=$$0$. When $V$ is further increased such that $V>V^\textrm{bulk}_c\approx 0.45t$, the bulk AF will be induced and the whole system breaks TR symmetry. The global quantum phase diagram is summarized in \Fig{fig1}(b). There is a line of edge QCP and a separated line of bulk QCP. In the intermediate range, the TR symmetry breaking occurs only on the edge while the bulk is still TR invariant.

\begin{figure}[t]
\subfigure[]{\includegraphics[height=3.2cm]{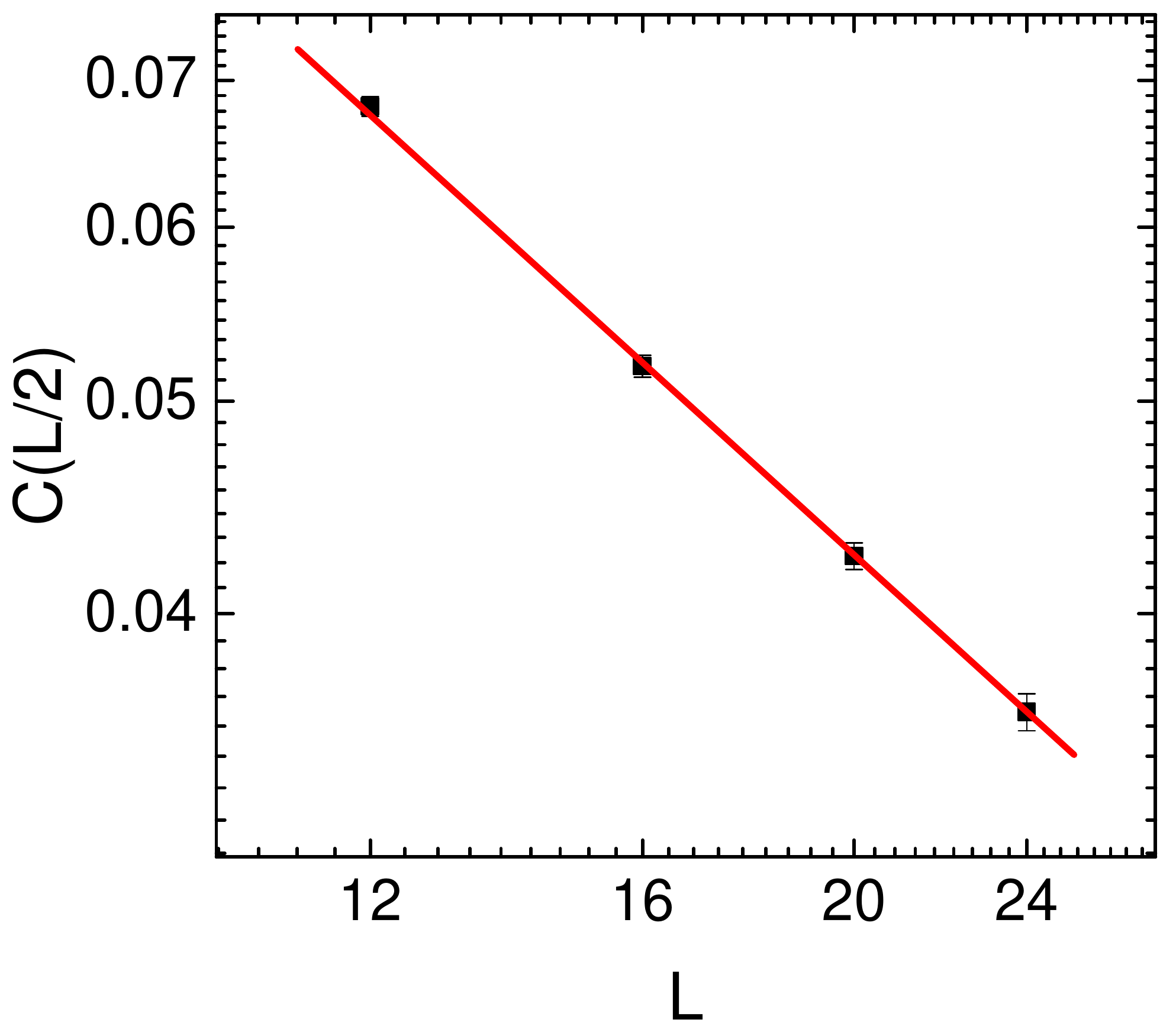}}~~~
\subfigure[]{\includegraphics[height=3.2cm]{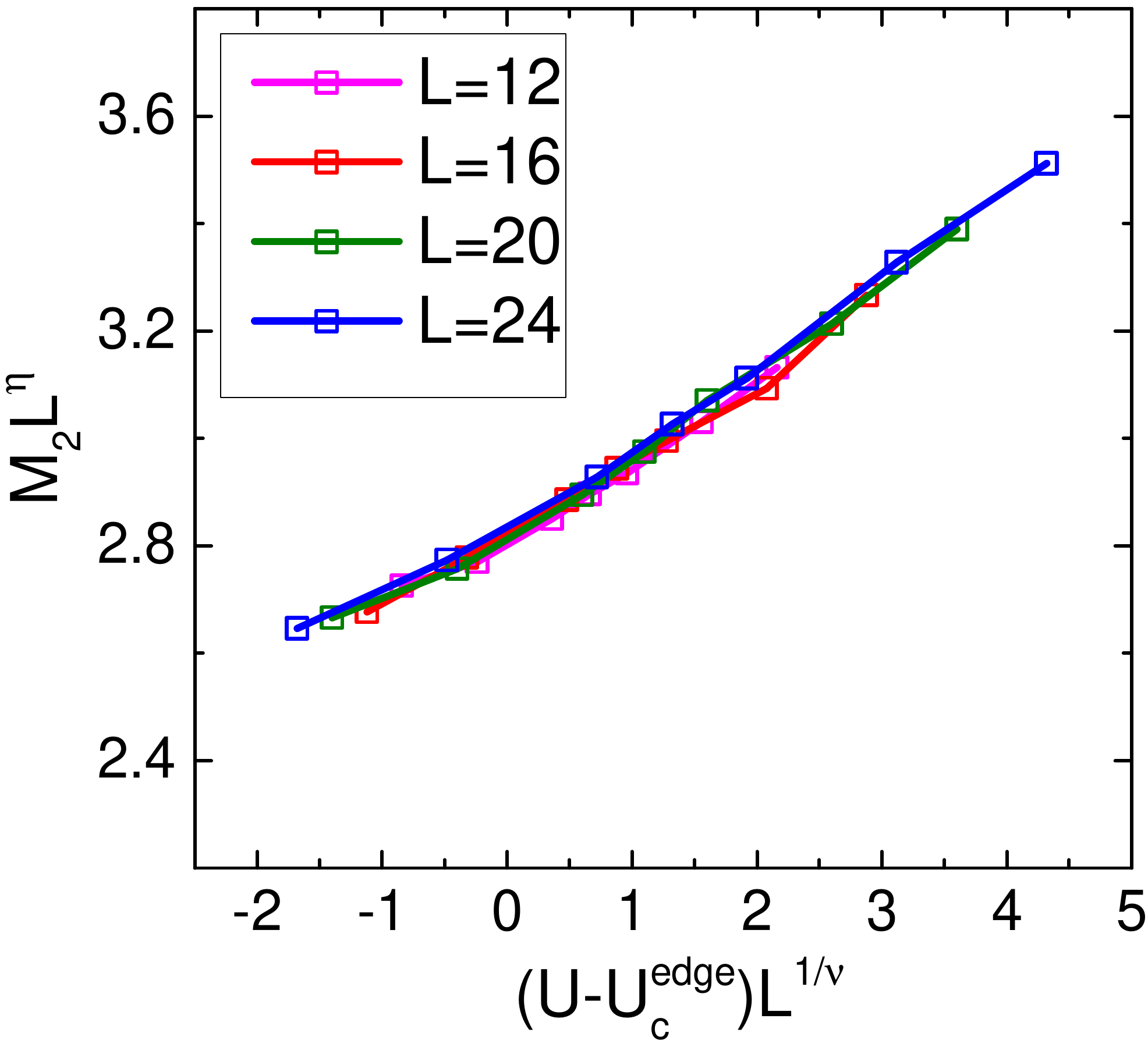}}							
\caption{The QMC results of the edge QCP at $V=0.3$. (a) The log-log plot of spin-spin correlation function at largest distance $C(\frac{L}{2})$ as the function of linear systems size $L$: $\eta = 0.94 \pm0.02$.  (b) Data collapse of magnetic structure factor at different values of $U$ and different systems size $L$: $\nu \!=\! \infty$. }
\label{fig3}
\end{figure}

\textbf{Edge quantum criticality:} It would be interesting to analyze the edge quantum critical behaviors \cite{Subirbook} in topological phases to study exotic phenomena such as the emergence of supersymmetry \cite{Ashvin2014, Ponte2014,Li2016arxiv,Jian2017prl}. The EQCP can be driven by either $U$ or $V$. Here, we consider a weak but finite spin-flip interaction $V=0.3t$ and vary $U$ in the QMC simulations to study the EQCP. Similarly, we use RG-invariant quantity $R^\textrm{edge}_c$ to identify the EQCP. The results of our sign-free MQMC simulation on $R^\textrm{edge}_c$ (see the SM for details) clearly show that the edge quantum phase transition occurs at $U^\textrm{edge}_c \approx 1.72t$. Moreover, we computed the evolution of the single-particle gap as the function of $U$ (see the SM for details). When the TR symmetry is spontaneously broken at $U>U^\textrm{edge}_c$, the helical edge state becomes gapped. Close to the $U^\textrm{edge}_c$, the gap (equivalently the inverse correlation length) scales as $\xi^{-1}$$\sim$$\Delta$$\sim$$\exp[-\frac{A}{\sqrt{U-U^\textrm{edge}_c}}]$, where $A$ is some constant.

We then investigate the quantum critical behaviours of this EQCP. When $U>U^\textrm{edge}_c$, the edge develops a finite magnetic order breaking the TR symmetry. The transition should belong to the universality class of Kosterlitz-Thouless transitions in 1+1 dimensions \cite{Fradkinbook}. By fitting the correlation function of magnetic order at the EQCP: $C(\frac{L}2)\sim L^{-\eta}$, we obtain the anomalous dimension of order parameter bosons: $\eta=0.94\pm0.03$, as shown in \Fig{fig3}(a). According to the bosonization analysis, the correlation length critical exponent $\nu$ of this EQCP should be infinite. In order to verify it, we perform data collapse of the edge magnetic structure factor $M_2 = \frac{1}{L}\sum_{r}C(r)$ by the scaling function $M_2 L^\eta = F(L^{1/\nu}(U-U^\textrm{edge}_c))$, $F$ is an unknown function. When $1/\nu = 0$ and $\eta = 0.94$, various points $(M_2 L^\eta, L^{1/\nu}(U-U^\textrm{edge}_c))$ of different $U$ around $U^\textrm{edge}_c$ and different $L$ can collapse to a single curve, as shown in \Fig{fig3}(b). Consequently, we verified that the EQCP belongs to the Kosterlitz-Thouless universality in 1+1 dimensions with $\eta=0.94\pm0.03$ and $\nu=\infty$.

\begin{figure}[t]
\subfigure[]{\includegraphics[height=2.8cm]{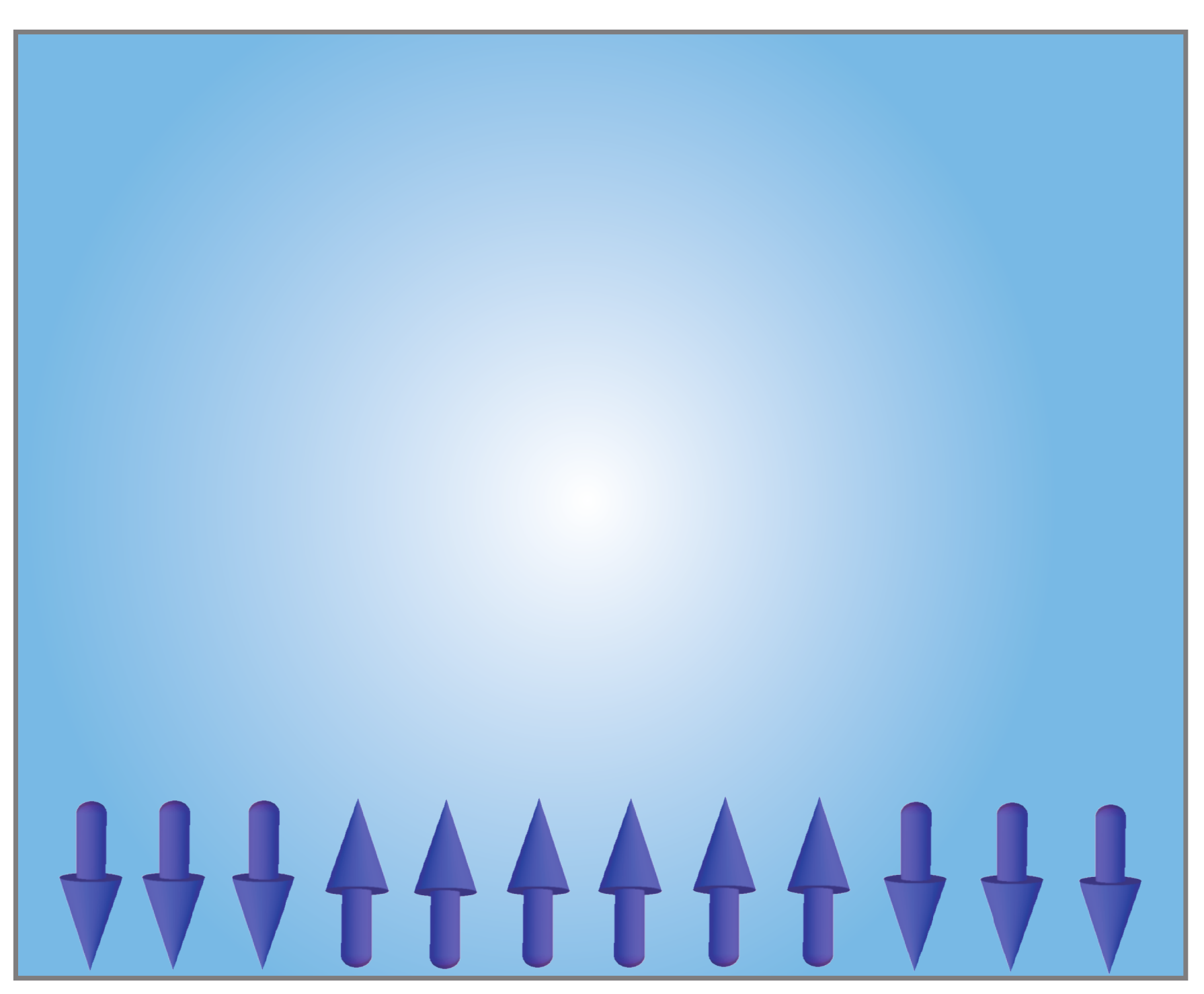}}
\subfigure[]{\includegraphics[height=2.8cm]{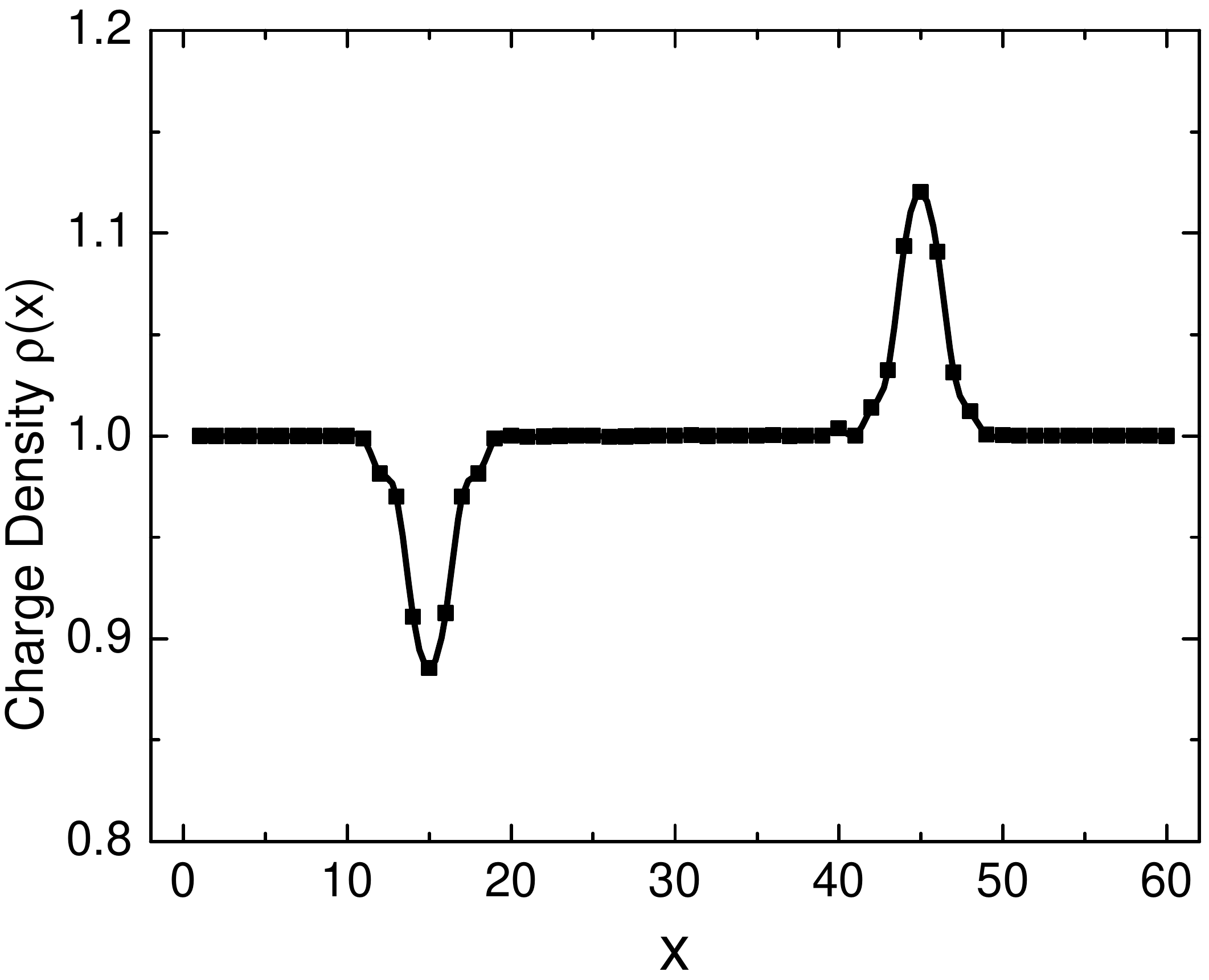}}
\caption{QMC results of distribution of electron density at boundary. Small Zeeman field $M_z$=0.05 is used to generate magnetic domain. The positions of magnetic domain walls are $x$=15 and $x$=45. Fractional charge with $\pm e/2$ are found at two mangetic domain walls.  }
\label{fig4}
\end{figure}

\textbf{Fractionalized charge:}
The edge magnetic ordering can render some exotic physics because the helical edge states of 2D TI has only half degrees of freedom of a one-dimensional system of spin-1/2 electrons respecting the TR symmetry. For instance, it was predicted theoretically that a fractionalized charge of $e/2$ can be induced at the magnetic domain wall in the helical edge states of a 2D TI system \cite{Qi20082}, which is a realization of the Jackiw-Rebbi mass soliton of the 1+1 Dirac theory \cite{Jackiw1976}. To verify the domain-wall fractionalized charge, an infinitesimal external magnetic field polarized in opposite directions in magnetic ordered phase to create a magnetic domain wall, as shown in \Fig{fig4}(a).

We set $V=0.3t$ and $U=4.0t>U^\textrm{edge}_c$ such that the edge is magnetically ordered while the bulk still respects the TR symmetry. The lattice sizes we choose are $L_x=60$ and $L_y=6$. Two magnetic edge domain walls are generated at $x$=15 and $x$=45, as shown in \Fig{fig4}(a), by imposing a weak external magnetic field $B_z = \pm 0.05$. We compute distribution of electron density $\rho(x)$ on the edge while the average electron density per site is $\bar\rho=1$. The excess charge localized at the magnetic domain wall is given by $n_\textrm{dw}=\sum_x \rho(x)-\bar \rho$, where $x$ is summed only around the domain wall. Our QMC results show that $n_\textrm{dw}\approx e/2$, as shown in \Fig{fig4}(b). This provides robust numerical evidences of the emergent fractionalized charge at magnetic domain wall in the \textit{interacting} helical edge states with spontaneous TR symmetry breaking. Such magnetic domain walls carrying fractionalized $e/2$ charge can be potentially measured in future experiments.

\textbf{Discussions and conclusions:}
We now discuss the implication of our results to recent experiments done in the 2D TI InAs/GaSb QW \cite{Du2015a,Du2015b}. Transport experiments in Ref. \cite{Du2015b} reported the temperature dependence of its edge conductance: $G_{xx}\sim T^{0.32}$ for sufficiently low temperature, which indicates that its edge Luttinger parameter is $K\approx 0.22$ assuming that the conductance is dominated by instantons at an impurity with fractionalized charges $e/2$ \cite{Joseph2009}. For $K<1/2$, the helical edge states can also be gapped by spontaneously breaking the TR symmetry on edges when the two-particle backscattering is allowed (namely when the Fermi level is tuned to near the Dirac point of the helical edge states). At finite temperature, charge transport could have contributions from magnetic domain wall with fractionalized charge of $e/2$, as shown above. It would be interesting to study transport properties of helical Luttinger liquids \cite{Oreg2012,Crepin2012,Schmidt2012prl,Mirlin2014,Glazman2016prb,Ziani2017np} with two-particle backscattering in the region of $K<\frac12$ by sign-free QMC simulations, which will be deferred to future works. As the helical edge states coupled with superconductors and with magnetism can support Majorana zero modes \cite{Hasan-Kane-10,Qi-Zhang-11}, it would also be interesting to study the effect of strong interactions on Majorana zero modes \cite{Kane2014prl,Schmidt2015prb}.

In conclusion, we have proposed the KMHR model to describe 2D interacting TIs allowing two-particle spin-flipping backscattering. It is the first two dimensional model with two-particle backscattering which can be simulated by sign-free QMC using the Majorana representation. Our large-scale QMC simulations of this model have shown that the helical edge states are robust under weak Rashba spin-flipping interaction when the Hubbard (or Coulomb) repulsion is not too strong. When Hubbard repulsion is strong enough, the gapless helical edge states are unstable again even infinitesimal two-particle backscattering by spontaneously breaking the TR symmetry and forming magnetic order at edges. The critical behaviors of this EQCP has also been obtained. Our work may provide a promising new direction to study the boundary stability and quantum criticality in topological phases of matter by non-perturbative approaches.

{\it Acknowledgements}: We sincerely thank Yi-Fan Jiang, Dung-Hai Lee and Shou-Cheng Zhang for helpful discussions. This work is supported in part by the NSFC under Grant No. 11474175 (ZXL and HY) and by the MOST of China under Grant No. 2016YFA0301001 (HY).

\begin{widetext}
\section{Supplementary Materials}

\renewcommand{\theequation}{S\arabic{equation}}
\setcounter{equation}{0}
\renewcommand{\thefigure}{S\arabic{figure}}
\setcounter{figure}{0}

\subsection{I. Proof of Sign-problem-free of interacting topological insulator model in Majorana representation}

To prove the absence of sign-problem in QMC simulations of the KMHR model in Eq. (1) of the main text, we introduce the Majorana representation of spin-1/2 electrons: $c_{i\sigma} = \frac{1}{2}(\gamma^1_{i\sigma} + i \gamma^2_{i\sigma})$, $c^\dagger_{i\sigma} = \frac{1}{2}(\gamma^1_{i\sigma} - i\gamma^2_{i\sigma})$, where $\gamma^\tau_{i\sigma}$ are Majorana fermions operators with $\tau=1,2$ representing Majorana index and $\sigma=\uparrow,\downarrow$ representing spin index. First, we perform the particle-hole transformation on $\sigma=\downarrow$ electrons: $c_{i\downarrow} \rightarrow (-1)^i c_{i\downarrow}^\dagger$. Under this transformation, the NN hopping term and NNN Kane-Mele SOC term are invariant. The Hubbard interaction term changes sign and Rashba interaction becomes $V(c^\dagger_{i\uparrow}c^\dagger_{i\downarrow}c^\dagger_{j\downarrow}c^\dagger_{j\uparrow} + h.c)$.  After the particle-hole transformation, the Hamiltonian in Eq. (1) can be rewritten in Majorana representation as:
\bea
H &=& H_0 + H_I \nonumber\\
  &=& \sum_{\avg{ij}} - \frac{t}{2} \gamma^T_i \sigma^0 \tau^t \gamma_j + \sum_{\avg{\avg{ij}}}  \frac{ i \lambda \nu_{ij}}{2} \gamma^T_i  \sigma^z\tau^0 \gamma_j - \sum_i \frac{U}{4} (i\gamma^1_{i\uparrow} \gamma^2_{i\uparrow})  (i \gamma^1_{i\downarrow}\gamma^2_{i\downarrow}) + \sum_{\avg{ij}} \frac{V}{32} \sum_{\alpha}^4 [i \gamma^T_{i} B_a \gamma_j]^2
\eea
where $\gamma^T_{i} \equiv (\gamma^1_{i\uparrow},\gamma^2_{i\uparrow},\gamma^1_{i\downarrow},\gamma^2_{i\downarrow})$, $B_1 = \sigma^z\tau^z, B_2= i \sigma^0\tau^z, B_3 = \sigma^0\tau^x, B_4 = i\sigma^z\tau^x$. Upon Trotter decomposition and usual Hubbard-Stratonovich transformations, the decoupled Hamiltonian at imaginary time $\tau$ can be written in Majorana representation as:
\bea\label{decoupledTI}
\hat{h} = \sum_{\avg{ij}} -\frac{t \Delta_\tau}{2} \gamma^T_i \sigma^0 \tau^t \gamma_j + \sum_{\avg{\avg{ij}}} \frac{ i \lambda \nu_{ij}\Delta_\tau}{2} \gamma^T_i i \sigma^z\tau^0 \gamma_j + \sum_{i} \lambda_U \varphi_i \gamma^T_i \sigma^0 \tau^y \gamma_i + \sum_{\avg{ij},\alpha} \lambda_V \phi^{\alpha}_{ij} \gamma^T_i B_{\alpha} \gamma_j
\eea
where $\Delta_\tau$ is imaginary time slice of the Trotter decomposition,  $\varphi_i$ are imaginary-time dependent auxiliary fields on site $i$ and $\phi^\alpha_{ij}$ are imaginary-time dependent auxiliary fields on bond $\avg{ij}$. The decoupled Hamiltonian in \Eq{decoupledTI} possesses two anti-commuting Majorana-time-reversal symmetries: $T^- = i \sigma^y \tau^x K$ and $T^+ = \sigma^x \tau^x K$. According to the Majorana TRS principle for sign-problem-free QMC, it belongs to Majorana class and is then sign-problem-free.

\subsection{II. Details of the projector Majorana QMC simulations}
We use projector QMC in the Majorana representation to investigate the ground state properties of the 2D interacting topological insulator described by the Hamiltonian in Eq. (1). In the projector QMC, the expectation value of an observable $O$ in the ground state can be evaluated as:
$\frac{ \bra{\psi_{0}} O \ket{\psi_0}}{ \avg{\psi_{0} \mid \psi_{0} } }   = \lim_{\Theta\rightarrow \infty} \frac{ \bra{\psi_T}e^{-\Theta H } O e^{-\Theta H} \ket{\psi_T}}{ \bra{\psi_T} e^{-2\Theta H} \ket{\psi_T}}$,
where $ \psi_0 $ is the true ground state wave function and $ \ket{\psi_T}$ is a trial wave function which should have a finite overlap with the true ground state wave function. Note that $\Theta$ is not the inverse temperature but the projection parameter. Although $\Theta\to\infty$ is needed to reach the exact ground state, in numerically calculations a sufficient large $\Theta$ works for practical purposes of obtaining physical quantities with required accuracy. Because of the absence of sign-problem, we can perform large-scale QMC simulations with large system sizes and sufficiently large $\Theta$. In the study of the bulk quantum phase transition, we use periodic boundary condition. In the study of edge quantum phase transition and edge Luttinger parameters, we use the periodic boundary condition in the $x$ direction but open boundary condition in the $y$ direction. In our QMC simulations, the imaginary-time projection parameter is $\Theta = 60/t$ for the systems with torus boundary conditions. In the cases of cylinder boundary conditions, most systems are computed using $\Theta = 75/t$ and some systems with large systems size or near critical points are computed using $\Theta = 100/t$. We have checked that all the results stay nearly the same when larger $\Theta$ are used, which ensures desired convergence to the limit of $\Theta\to\infty$. We set $\Delta_\tau =0.05/t$ and the results do not change if we use smaller $\Delta_\tau$.

\subsection{III. Details of the RG-invariant quantity $R^\textrm{edge}_c$ and single-particle gap }

We use the RG-invariant ratio $R^\textrm{edge}_c$ in terms of second-momentum correlation length of edge AF magnetic order to identify the EQCP. The second-momentum correlation length of edge AF magnetic order is defined as
\bea
\xi = \frac{L_x}{2\pi} \sqrt{\frac{S(\vec{0})}{S(\delta\vec{k})}-1},
\eea
where $\delta \vec {k}$ is the minimum lattice momentum internal for systems of size $L_x$. Here $S(\vec{k})$ is the structure factor of magnetic order at edge: $S(\vec{k}) = \frac{1}{L_x}\sum_{r} C(r) e^{i \vec{k}\cdot \vec{r}}$.  The ratio $R^\textrm{edge}_c$ is defined as $R^\textrm{edge}_c = \frac{\xi}{L}$. At the putative EQCP, $R^\textrm{edge}_c$ is RG-invariant, such that it should cross at the same point for different system sizes. In the disordered phase, the ratio $R^\textrm{edge}_c = \frac{\xi}{L}$ should decrease as the system size is increased. The trend is opposite in the ordered phase. Thus, this RG-invariant ratio is a powerful tool to accurate identify the QCP. Here, we fix the strength of spin-flip Rashba interaction $V=0.3t$ and vary $U$ to study the EQCP. The RG-invariant ratio $R^\textrm{edge}_c$ is evaluated by sign-free QMC. The results (shown in \Fig{figs1}) explicitly show that the EQCP occurs at $U \approx 1.72 t$.

When the TR symmetry is spontaneously broken at edge by form magnetic ordering, the helical edge state becomes gapped. The single-particle gap can be obtained in QMC through measuring the tails of the imaginary-time displaced Green's function:
\bea
G^f_k(\tau) = \sum_{\sigma=\uparrow,\downarrow}\avg{c^\dagger_{k\sigma}(\tau)c_{k\sigma}(0)},
\eea
where $\tau$ represents imaginary time, $c^\dagger_{k\sigma}(\tau) = e^{\tau H} c_{k\sigma}^\dagger e^{-\tau H}$, and $k$ is the edge momentum.
Here $k$ is the momentum in $x$ direction and $c^\dagger_k = \frac{1}{L_x}\sum_{x=1}^{L_x} c^\dagger_{(x,y=L_y)} e^{ik x}$. The single-particle gap $\Delta_{sp}$ corresponds to the single-particle excitation energy at $k=0$, which can be obtained from $G^f_{k=0}(\tau) \propto e^{-\tau \Delta_{sp}}$ when $\tau$ is large enough. We evaluate the single-particle gap as the function of $U$. The result clearly shows that when TR symmetry is spontaneously broken at $U>U_c^{\textrm{edge}}$, the single-particle gap in the helical edge states is opened.

\begin{figure*}[t]
\subfigure[]{\includegraphics[height=4.5cm]{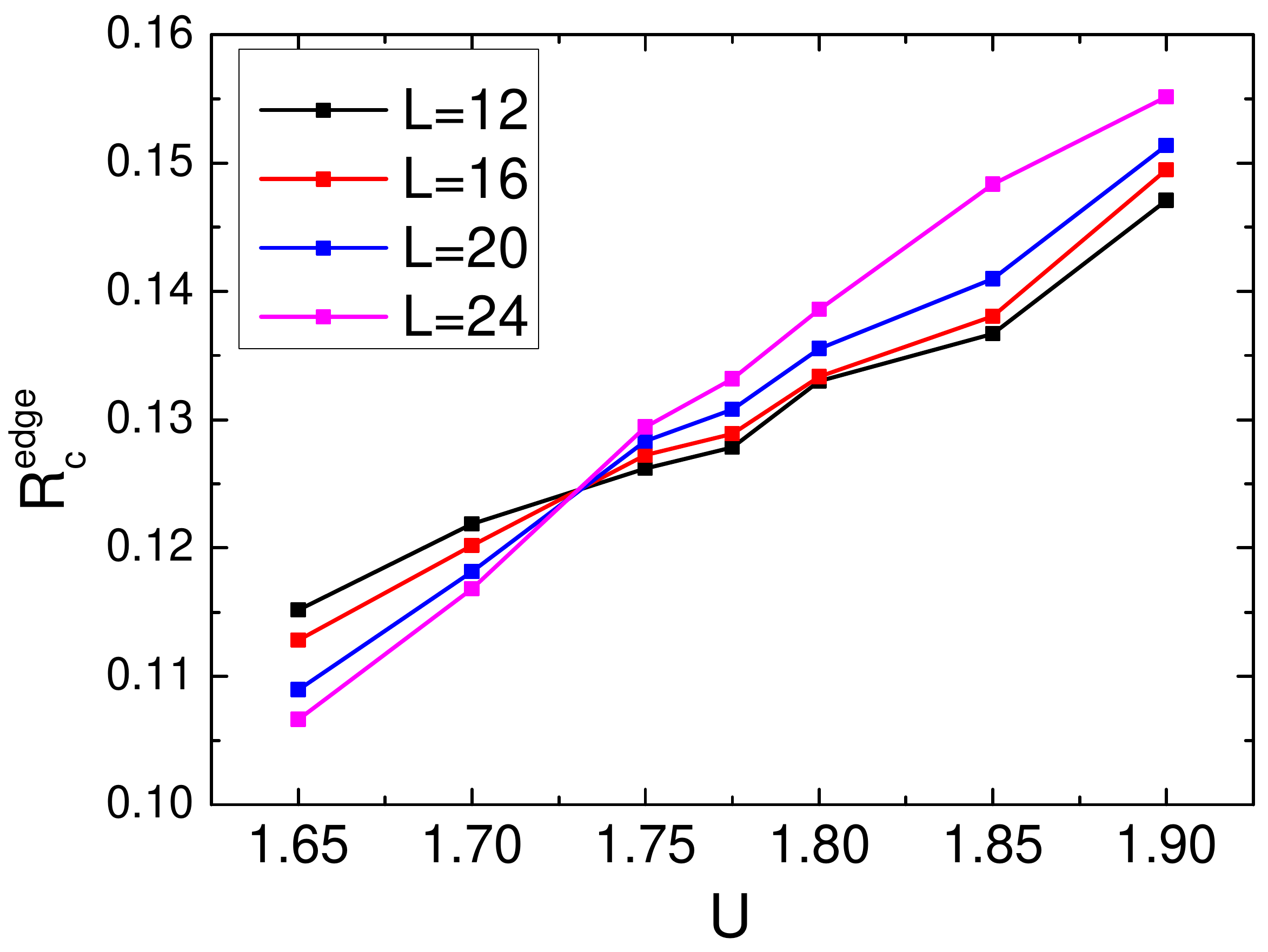}}
\subfigure[]{\includegraphics[height=4.5cm]{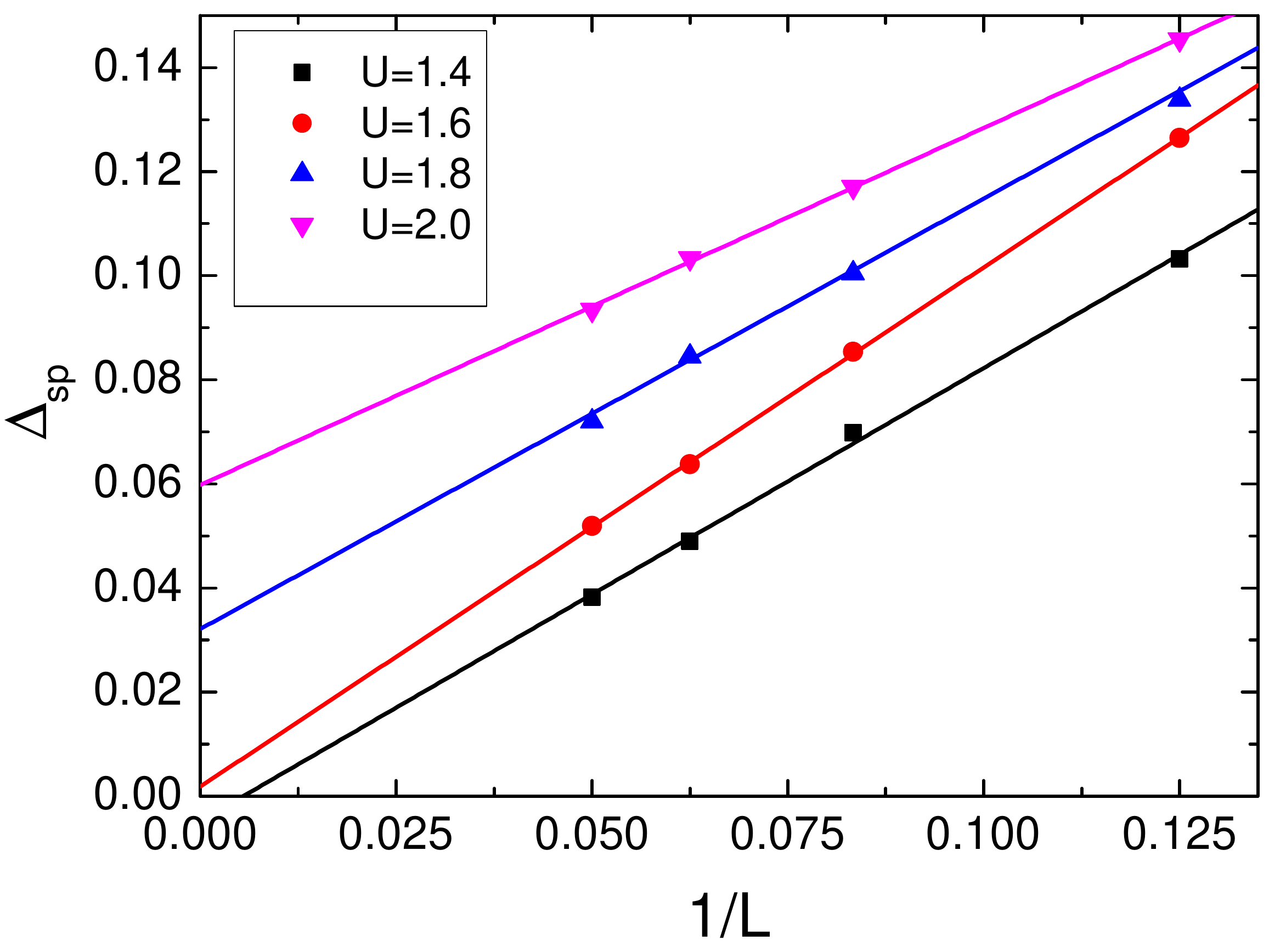}}							
\caption{The QMC results of the edge QCP at $V=0.3$:
(a)RG-invariant quantity $R^{\textrm{edge}}_{c}$ as the function of $U$. The crossing point explicitly shows that the transition point is $U \approx 1.72t$.  (b) Single-particle gap in edge helical state as the function of $U$. When hubbard interaction $U$ is larger than $U_c^{\textrm{edge}}$, single-particle gap is opened at edge.  }
\label{figs1}
\end{figure*}

\subsection{IV. Numerical results of bulk AF quantum phase transition}

When the Hubbard interaction $U$ or spin-flip Rashba interaction $V$ is strong enough, the spontaneous TR symmetry breaking should also occurs in the bulk. We use similar techniques including the RG-invariant ratio and data collapse to study the bulk quantum phase transition and analyze the quantum critical point. We fix $V=0.3t$ and vary the value of $U$. Similar to the case of EQCP, we measure RG-invariant ratio $R^\textrm{bulk}_c$ of the bulk AF magnetic order to determine the bulk quantum phase transition point. From the crossing point of $R^\textrm{bulk}_c$ in different systems sizes, we identify the bulk phase transition point $U_c^{\textrm{bulk}}=5.15t$, which is much larger than the edge one $U_c^{\textrm{edge}}=1.72 t$. It clearly indicates that the edge phase transition occurs when bulk is still disordered. We also employed the data collapse technique to study the quantum critical behaviour of this bulk AF phase transition. The structure factor of bulk AF order $M_2$ at different values of $U$ and for different systems should collapse to a single curve  $M_2 L^{1+\eta} = F(L^{1/\nu}(U-U^\textrm{bulk}_c))$. Consequently, we verified that the bulk AF quantum phase transition belongs to the Ising universality in 2+1 dimensions with $1+\eta = 1.01\pm 0.02$ and $\nu = 0.63\pm 0.03$.

\begin{figure*}[t]
\subfigure[]{\includegraphics[height=4.5cm]{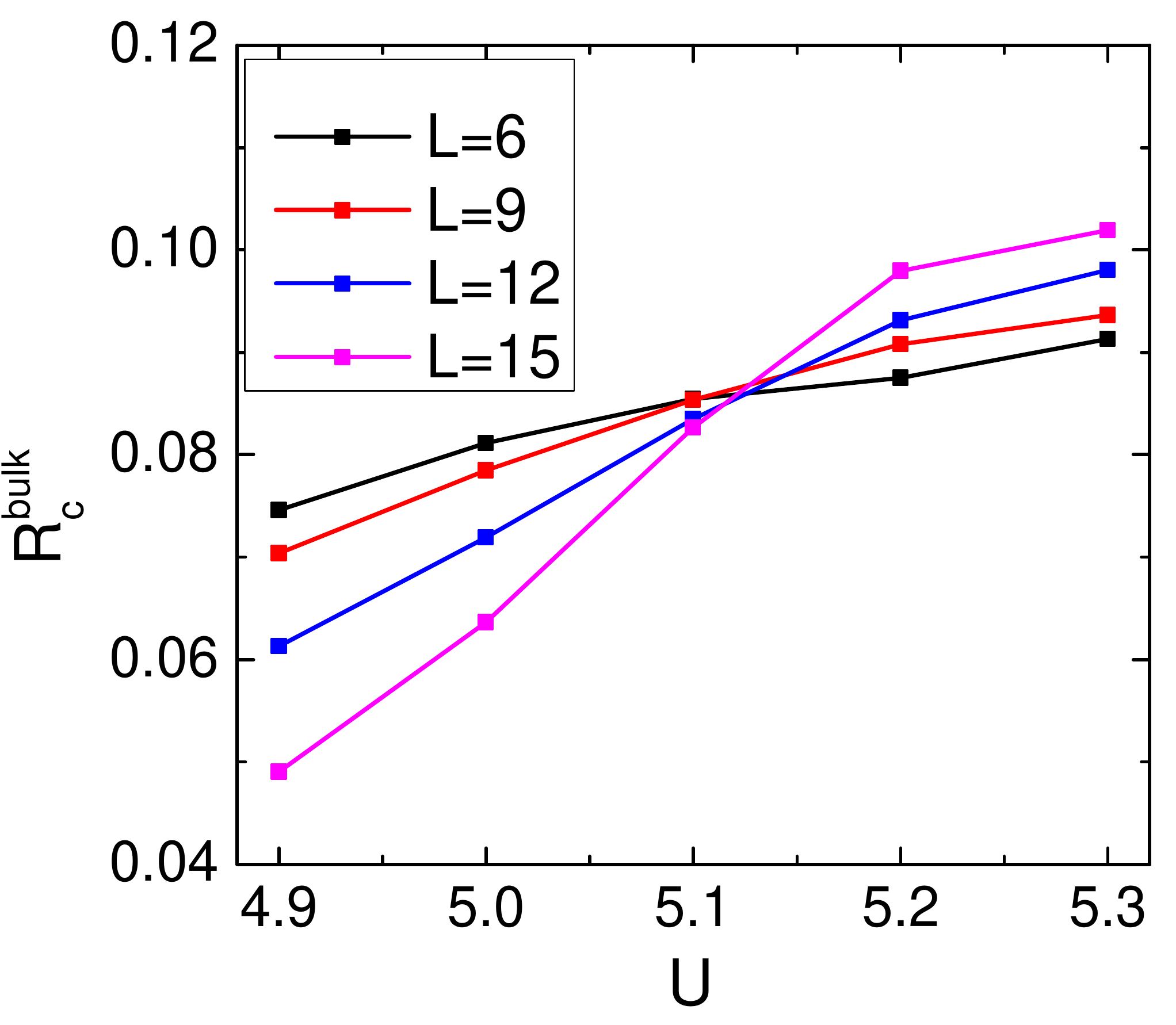}}
\subfigure[]{\includegraphics[height=4.5cm]{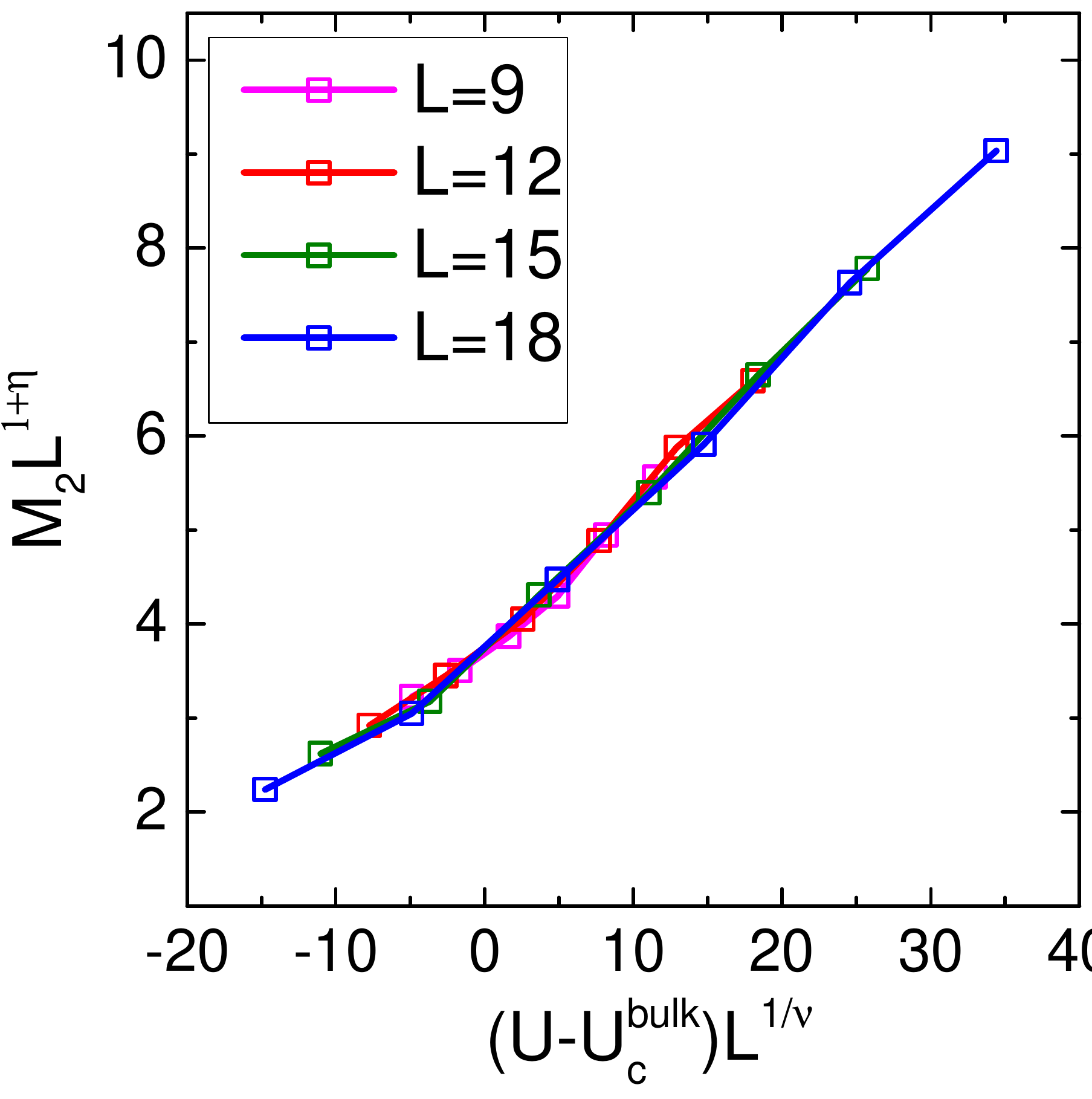}}							
\caption{The QMC simulations of the AF quantum phase transition in the bulk. (a) The RG-invariant ratio $R^{\textrm{bulk}}_{c}$ as a function of $U$. The crossing point explicitly shows that the bulk phase transition point $U_c^{\textrm{bulk}} \approx 5.15 t$. (b) From the data collapse of magnetic structure factor in the bulk at different values of $U$ and different systems size $L$, we obtain $1+\eta = 1.01\pm 0.02$ and $\nu = 0.63\pm 0.03$. }
\label{figs3}
\end{figure*}

\end{widetext}

\end{document}